\documentclass{PoS}

\title{\begin{center}
BL Lac objects with optical jets:\\ PKS 2201+044, 3C 371 and PKS 0521-365.
\end{center}}

\ShortTitle{BL Lac objects with optical jets.}

\author{\speaker E. Liuzzo
         \\
        Istituto di Radioastronomia-INAF-Bologna (Italy)\\
        E-mail: \email{liuzzo@ira.inaf.it}}
\author{ R. Falomo\\
        Osservatorio Astronomico di Padova-INAF (Italy)\\
}
\author{ A. Treves \\
        Universit\`a dell'Insubria (Como-Italy), associated to INAF and INFN\\
}

\abstract{We investigate the properties of the three BL Lac objects,  PKS 2201+044, 3C 371 and PKS 0521-365, that exhibit prominent optical jets. We present high resolution near-IR images of the jet of the first two, obtained with an innovative adaptive-optics system (MAD) at ESO VLT telescope. Comparison of the jet in the optical, radio, NIR and X-ray bands reveals strong similarities in the morphology. A common property of these sources is the presence of broad emission lines in their optical spectra at variance with the typical featureless spectrum of the nearby BL Lac objects. Despite some resemblances (e.g. in the radio type), significant differences (e.g. in the central black hole masses and radio structures) with radio-loud NLS1s are found.}

\FullConference{Narrow-Line Seyfert 1 Galaxies and their place in the Universe - NLS1,\\
		April 04-06, 2011\\
		Milan Italy}

\begin{document}

\section{Introduction}

Radio loud (RL) Active Galactic Nuclei (AGN), in the contrary to their radio quiet (RQ) counterparts, show prominent jets mainly observable in the radio band. Blazars, including BL Lac objects and flat-spectrum radio quasars (FSRQs), are an important class of RL AGNs in which jets are relativistic and beamed in the observing direction. This is the case for some nearby BL Lac sources characterized by weakness of lines, luminous, rapidly variable  non-thermal continuum, significant polarization, strong compact flat spectrum radio emission, and superluminal motion (e.g. \cite{ULRICH, GIROLETTI}) plus, in some cases, by strong gamma emission (\cite{ABDO}).

However, jets in BL Lacs are often difficult to resolve at high frequencies because of the close alignment with the line of sight. This is due to the fact that the detection of the jet at the different frequencies depends both on the limited angular resolution and on the short lifetime of the high energy electrons producing the non- thermal emission of the jet. 

The advent of high resolution imaging facilities such as HST and Chandra allowed systematic searches for angular resolved multiwavelength counterparts of radio jets in Blazars (\cite{SCARPA}) and also the detection of jets in radio-loud quasars at high redshift (\cite{SCHWARTZ, SAMBRUNA02, TAVECCHIO}).  Moreover, the use of adaptive optics assisted imaging on large telescopes permits to investigate from the ground the near-IR jet properties (\cite{HUTCHINGS, FALOMO09}).

PKS 0521-365, PKS 2201+044 and 3C 371 are up to now the only three classified BL Lac objects showing resolved optical jets. Near-IR observations (MAD) using an innovative adaptive optics (AO) camera of PKS 0521-365 and PKS 2201+044 were performed to investigate their morphological and photometric properties. We discuss the new observations together with the available data in radio, optical, and X-ray bands.
Here, we present our high resolution MAD near-IR data and our results on the comparative analysis of these three sources.

\section{MAD near-IR observations.}
We performed Ks-band observations of PKS 0521-365 (\cite{FALOMO09}) and PKS 2201+044 (\cite{LIUZZO}) using the European Southern
Observatory (ESO) Multi-Conjugate Adaptive Optics Demonstrator (MAD), mounted at UT4 (Melipal) of the Very
Large Telescope (VLT).
MAD ($57\hbox{$^{\prime\prime}$ }\times57\hbox{$^{ \prime\prime}$ }$ field of view and pixel size's detector of 0 $\!\!^{\prime\prime}$.028) is a prototype Multi Conjugate Adaptive optics (MCAO) system which aims to demonstrate
the feasibility of different MCAO reconstruction techniques in
the framework of the E-ELT concept and the 2nd Generation VLT Instruments (\cite{RAGAZZONI96, RAGAZZONI00}, see also \cite{MARCHETTI}).

\clearpage

\begin{figure}[h!]
  \centering
    \includegraphics[width=.4\textwidth]{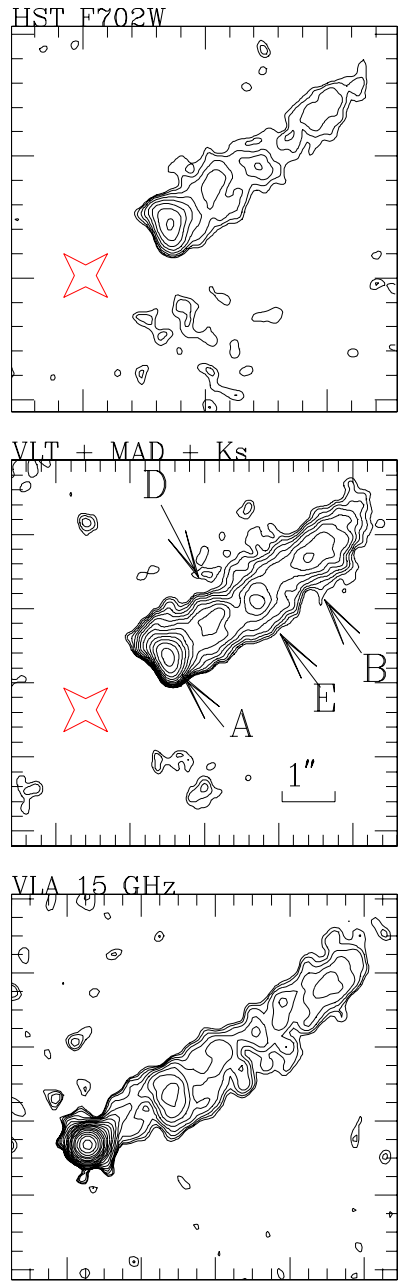}
  \caption{The contour plot of the jet of {\bf PKS 0521-365} observed by MAD in the Ks band (middle panel) compared with the image of the jet observed in  R band  by HST+WFPC2 (top panel) and the radio map at 15 GHz obtained by VLA (bottom panel).
The large star represents the position of the (subtracted) nucleus in both
optical and near-IR bands. In the maps, N is up and E is left. In the bottom panel, the radio jet is shown at the
same angular resolution as the near-IR data. The nuclear source has not been
subtracted. A collimated jet with the same position angle as the VLBI pc-scale
jet is present near the core. The jet becomes resolved
    transversely at less than 1$^{\prime\prime}$ from the core. See \cite{FALOMO09} for details.}
\end{figure}

\begin{figure} [h!]
\centering
\vspace{1cm}
\includegraphics[width=5.5 cm]{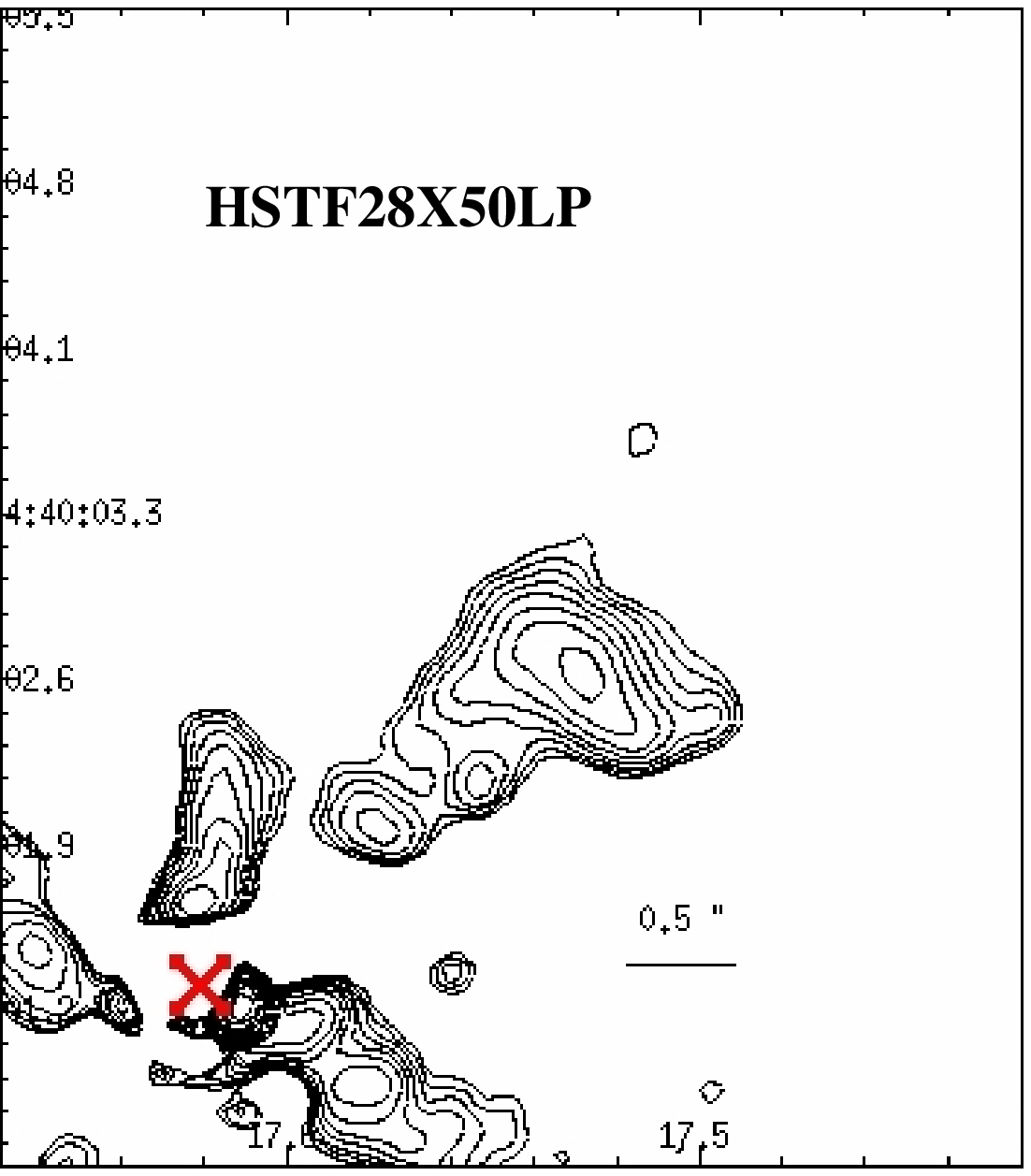} 
\hfill
\\
\includegraphics[width=5.5 cm]{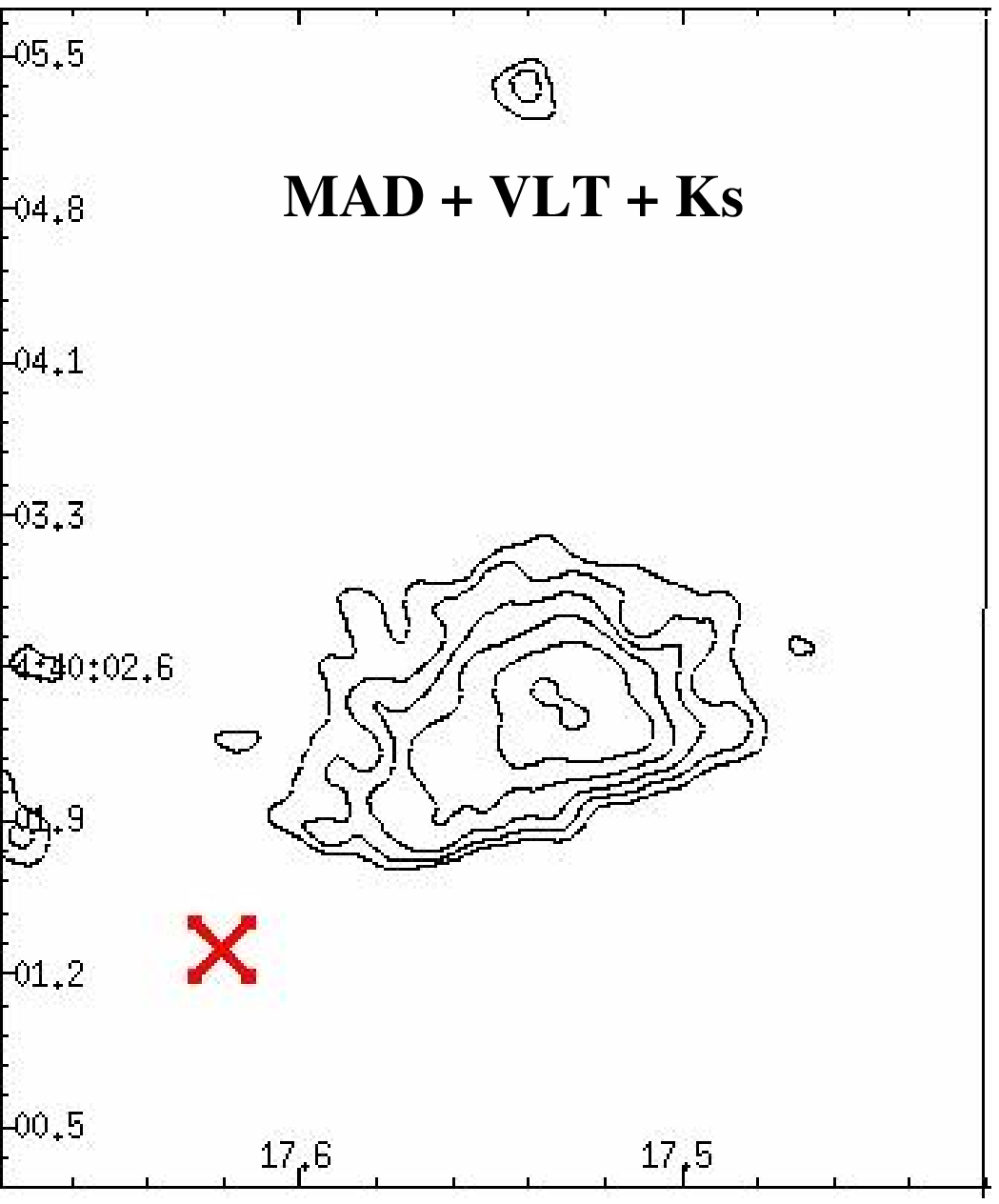} 
\hfill
\\
\includegraphics[width=5.5 cm]{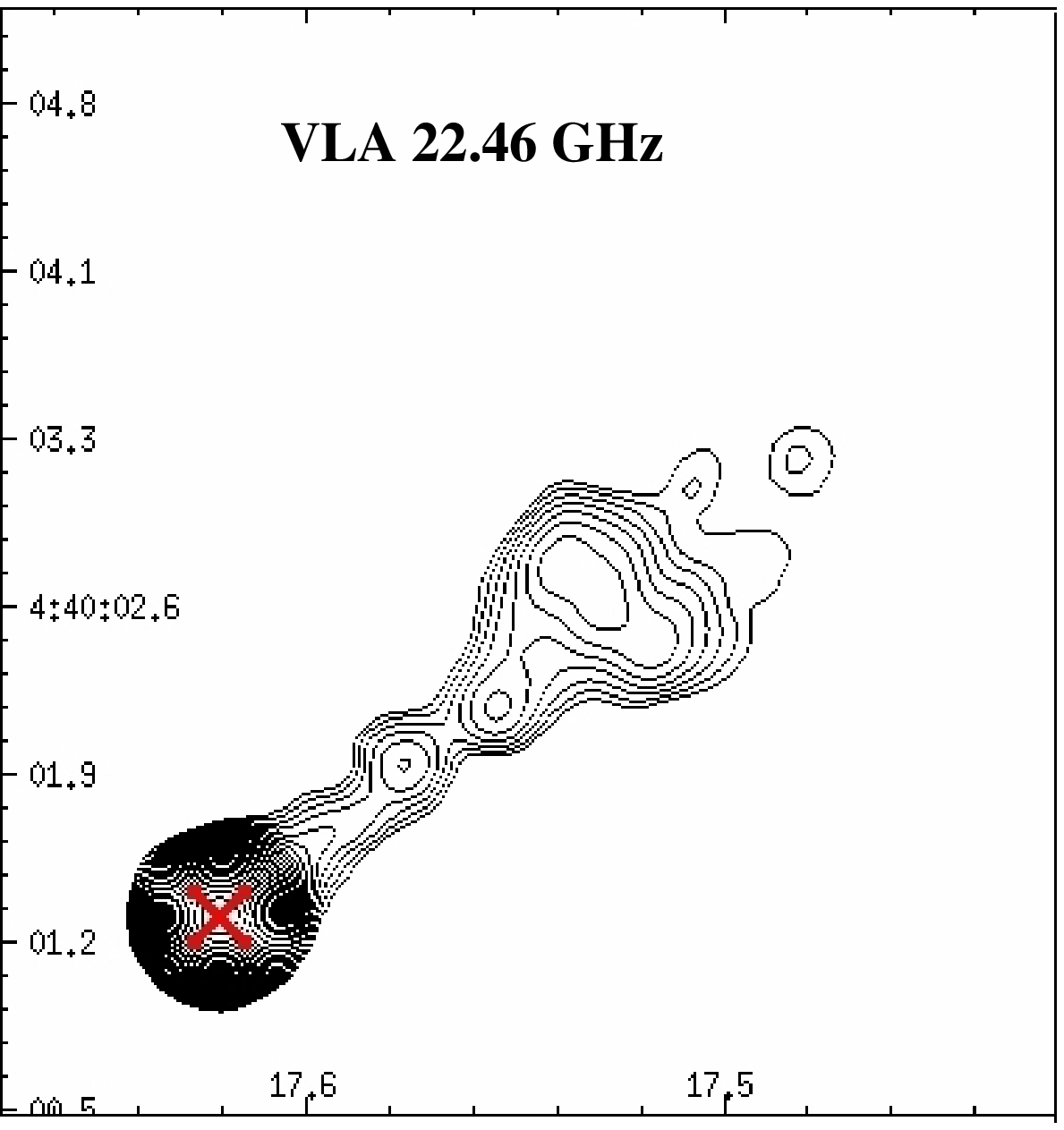} 
\caption{Contour plot of the jet emission of {\bf PKS 2201+044} observed by MAD in the Ks band (middle panel) compared with the image of the jet observed by HST (top panel) and the radio map at 22.46 GHz obtained by VLA (bottom panel). The red large cross represents the position of the (subtracted) nucleus in both
optical and near-IR bands. The resolution of the maps is $\sim$0.25 arcsec. Contour levels for the radio image are 0.2, 0.3, 0.4, 0.6, 0.8, 1, 2, 4, 8, 12, 15, 30, 50, 60, 100 and 200 mJy/beam.
The peak level is 0.28 Jy/beam. See \cite{LIUZZO} for details. }\label{rir}
 \end{figure}

\clearpage
\section{Results.} \label{pks}

Table 1 collects information on the properties of three sources. In Figs. 1-2 we report the morphology in optical, radio, NIR bands for PKS 0521-365 and PKS 2201+044. For 3C 371 refer to \cite{SAMBRUNA07}.  

\begin{table*}
\caption{Properties of PKS 2201+044, 3C 371 and PKS 0521-365.}\label{tabprop}
\scriptsize 
\begin{center}
\begin{tabular}{|c c c c c c c c c c|}
\hline
&&&&&&&&&\\
Object        & z    &  Galaxy & Spectral & Optical/UV   &  Radio & Radio      & Fermi            & $\delta$ & $\theta$ \\
              &      &   type  &  type    &  spectrum    &  type  & dimension & luminosity        &          &   \\
              &      &         &          &              &        &  (kpc)     & (10$^{44}$erg/s) &          &   \\
\hline
\hline
&&&&&&&&&\\
PKS 2201+044 & 0.027 & E & LBL & N +B                   & FRI & 162   & 0.2  & 2.1 & 29 \\   
&&&&&&&&&\\
\hline
&&&&&&&&&\\   
3C 371       & 0.051 & E & LBL & N + B                  & FRI & 70    & 1.92 & 3.5 & 17 \\
&&&&&&&&&\\
\hline
&&&&&&&&&\\    
PKS 0521-365 & 0.055 & E & LBL & N + B                  & FRI & 22    & 3.41 & 2.6 & 21 \\
&&&&&&&&&\\    
\hline  
\multicolumn{10}{l}{Col.1: name of the studied BL Lacs; Col. 2: BL Lac's redshift (z); Col. 3: BL Lac's galaxy type: E is for Elliptical galaxy type;}\\
\multicolumn{10}{l}{ Col.4: LBL indicates Low frequency BL Lac; Col.5: N+B is refered to the presence of Narrow and Broad emission lines;} \\
\multicolumn{10}{l}{ Col. 6: FRI is for Fanaroff-Riley radio type; Col.7: BL Lac's linear dimension in the radio band expressed in kpc; }\\
\multicolumn{10}{l}{Col. 8: Gamma luminosity as infered by Fermi detection in 10$^{44}$erg/s unit; Col. 9: $\delta$ is the Doppler factor estimated for the nucleus;  } \\
\multicolumn{10}{l}{Col. 10: $\theta$ is the viewing angle in degree.}\\
\end{tabular}
\end{center}
\end{table*}

In the following, we summarize the main results of our comparative analysis of these three BL Lacs:
\begin{itemize}
\item The properties of the jets of the three sources are remarkable similar. 
\item The jet morphology in optical, radio, NIR and X-ray bands has close resemblances with the presence of many knots (see Figs. 1-2). In particular, the brightest knot is observed in all bands from optical to X-rays. 
\item  The spectral radio index of the brightest knot is $\sim$0.7 for all the cases, which is typical of the emission due to the losses in the jet interaction with the surrounding medium. \\
The SED (Spectral Energy Distribution) of the  brightest knot is consistent with an emission dominated by a single synchrotron component.
\item The detection of resolved optical jets are consistent with estimates of the doppler factors and the viewing angles (\cite{GIROLETTI}).
\item The optical spectra of these BL Lacs exhibit weak narrow and broad emission line.  They are similar to that to BL Lac itself for which a Seyfert 1-like nucleus is reported (\cite{CORBETT}).
\end{itemize}

Finally, these BL Lacs have some properties (e.g. the source's spectral radio index and the FRI radio type) similar to those of radio-loud NLS1s with blazar-like characteristics (\cite{YUAN}). However, they show important differences with respect to the latters as they are not unresolved and not all compact on arcsecond scale (see Table 1, Figs. 1-2). Moreover, in the optical spectra, they show broad H$_{\beta}$ FWHM ($\sim$4500 km/s) and the absence of strong Fe II. On the other hand, central black holes estimates are significantly higher (Log(M$_{bh}$ /M$_{sol}$) $\sim$ 8.5, \cite{WOO}) while for NLS1s small black holes masses are expected (\cite{KOMOSSA}).

\section {Acknowledgments}
E.L.'s work was supported by contributions of European Union, Valle D'Aosta Region and the Italian Minister for Work and Welfare.
\\
\\

\end{document}